\documentclass[aps,prb,twocolumn,superscriptaddress,floatfix,letterpaper]{revtex4-1}

\pdfoutput=1

\usepackage{euscript}

\newcommand{\tAw}[3]{{#1}_{e_{#2#3} }^{v_{#2}v_{#3}}}

\newcommand{\tC}[5]{{#1}_{v_{#2}v_{#3}v_{#4}v_{#5};
\phi_{#2#3#5} \phi_{#3#4#5}
}^{
e_{#2#3} e_{#2#4} e_{#2#5} e_{#3#4} e_{#3#5} e_{#4#5};
\phi_{#2#3#4} \phi_{#2#4#5}
}}

\usepackage{amsthm}
\newtheoremstyle{wenthm}% name of the style to be used
  {3pt}% measure of space to leave above the theorem. E.g.: 3pt
  {3pt}% measure of space to leave below the theorem. E.g.: 3pt
  {\slshape}% name of font to use in the body of the theorem
  {}% measure of space to indent
  {\bfseries}% name of head font
  {:}% punctuation between head and body
  {.5em}% space after theorem head; " " = normal interword space
  {}% Manually specify head

\theoremstyle{wenthm}

\newtheorem{conj}{Conjecture}

\theoremstyle{definition}

\newtheorem{defn}{Definition}

\begin{document}

\begin{titlepage}

\title{Volume and Topological Invariants of Quantum Many-body Systems} 

\author{Xiao-Gang Wen}
\affiliation{Department of Physics, Massachusetts Institute of
Technology, Cambridge, Massachusetts 02139, USA}

\author{Zhenghan  Wang}
\affiliation{ Microsoft Station Q and Department of Mathematics, University of California, Santa Barbara, CA 93106, USA }

\begin{abstract} 
A gapped many-body system is described by a path integral on a space-time
lattice $C^{d+1}$, which gives rise to a partition function $Z(C^{d+1})$ if
$\partial C^{d+1} =\emptyset$, and a wave function $|\Psi\rangle$ on the
boundary if $\partial C^{d+1} \neq\emptyset$.  We show that $V = \text{log}
\sqrt{\langle\Psi|\Psi\rangle}$ satisfies the inclusion-exclusion property
$\frac{V(A\cup B)+V(A\cap B)}{V(A)+V(B)}=1$ and behaves like a volume of the
space-time lattice $C^{d+1}$ in large lattice limit (\ie thermodynamics limit).
This leads to a proposal that the vector $|\Psi\rangle$ is the quantum-volume
of the space-time lattice $C^{d+1}$.  The quantum-volume satisfies a quantum
additive property.  The violation of the inclusion-exclusion property by $V =
\text{log} \sqrt{\langle\Psi|\Psi\rangle}$ in the subleading term of
thermodynamics limit gives rise to topological invariants that characterize the
topological order in the system.  This is a systematic way to construct and
compute topological invariants from a generic path integral.  
For example, we
show how to use non-universal partition functions $Z(C^{2+1})$ on several
related space-time lattices $C^{2+1}$ to extract $(M_f)_{11}$ and
$\text{Tr}(M_f)$, where $M_f$ is a representation of the modular group
$SL(2,\mathbb{Z})$ -- a topological invariant that almost fully characterizes
the 2+1D topological orders.

\end{abstract}


\maketitle

\end{titlepage}

%{\small \setcounter{tocdepth}{1} \tableofcontents }

\noindent
\textbf{Introduction}: 
Recently, it was proposed that all force particles (the gauge bosons) and
matter particles (the fermions) may arise from entangled quantum information if
we assume the space to be an ocean of qubits
\cite{W0303a,LW0622,W1301,YBX1451,YX14124784}.  If the physical space is indeed
an entangled ocean of qubits, then it is natural to suspect that the
mathematical notion of continuous space (\ie the notion of manifold) may also
arise from entangled qubits that are discrete and algebraic in nature.  This
leads to a current very active research direction trying to view continuous
geometry as emergent from discrete algebra.  This point of view may lead to a
quantum theory of gravity \cite{M9831,L1067} -- a long-sought-after goal of
fundamental theoretical physics.  However, at the moment, we still do not know
how the metrics of a manifold, and Einstein equation that govern the dynamics
of metrics as the \emph{only low energy excitations}, can emerge from discrete
and entangled qubits.  (For the emergence of non-Einstein quantum gravity as
the only low energy dynamics, see \Ref{X0643,GW0600,GW1290}.) In this paper, we
will address a much simpler question: how the volume emerges from discrete and
entangled qubits. We would like to demonstrate that at least one geometric
quantity, the volume, can emerge from discrete algebra.  

It turns out that if we only have emergent volume, the associated space does
not have a sense of ``shape'' and its dynamics is not governed by Einstein's
theory of gravity, but by a different gravitational theory -- a topological
quantum field theory \cite{W8951,A8875}.  We may call this kind of gravity as
topological gravity.  There are many examples to demonstrate how various
topological gravity (\ie various topological quantum field theories) emerge
from entangled qubits (\ie entangled many-body systems).  The emergence of
topological quantum field theories from entangled many-body systems is well
studied in condensed matter physics under the name of topological order
\cite{W9039,WN9077}.  Thus, entangled
many-body systems can also give us topological gravity and a sense of volume
-- an emerging geometric property.

At the first sight, the issue of emergent volume appears to be trivial for
many-body systems, since every many-body system has a natural definition of
volume: the number of lattice sites.  However, this only works for
translation symmetric many-body system.  For many-body systems without
translation symmetry, it is not proper to define the volume as the number of
lattice sites.  Now we can state the main issue that we try to address in this
paper: \emph{how to define the notion of volume for a non-translation symmetric
many-body systems on lattice}?

We find that if a quantum many-body system is in a topologically ordered phase
(or more precisely, a gapped quantum liquid state), then the notion of volume
can be defined even without translation symmetry.  However, the volume that
directly arise from the many-body system is not exactly the volume in the
familiar classical sense.  We will call the new notion of volume as quantum
volume.  Unlike classical volume which is a positive real number, a quantum
volume is not a real number, but a vector in a Hilbert space.  

From the quantum volume of a many-body system, we may define an emergent
classical volume as the norm of the quantum volume (\ie the norm of the
vector).  We find that such classical volume does not satisfy the classical
volume axioms exactly. However, in the large system size limit (the thermodynamical limit), 
the leading term of the classical volume does satisfy the
classical volume axioms.

We also find that the finite subleading terms that violate the classical volume
axioms vanishes for many-body states with trivial topological order (\ie for
product states).  So non-vanishing subleading terms imply a non-trivial
topological order.  In fact, those finite subleading terms are topological
invariants that characterize the underlying topological order.

This is very similar to entanglement entropy: the leading term of  entanglement
entropy can be used to define the total area of the interface, while the finite
subleading term -- the topological entanglement entropy -- is a topological
invariant that characterize the underlying topological order
\cite{KP0604,LW0605}.  We speculate that the two could be related by some
generalization of the Fubini's theorem.

~

\noindent
\textbf{Volume in quantum many-body system}:
To define a many-body system through a space-time path integral, we first
triangulate the $d+1$-dimensional space-time to obtain a simplicial complex
$C_N$ with $N$ vertices. The degrees of freedom of our lattice model live on
the vertices (denoted by $v_i$ where $i$ labels the vertices), on the edges
(denoted by $e_{ij}$ where $\<ij\>$ labels the edges), \etc.  The action
amplitude $\ee^{S_\text{cell}}$ for an $n$-cell $(ij \cdots k)$ is complex
function of $v_i$, $e_{ij},\cdots$ on the cell: $T_{ij \cdots
k}(\{v_i\},\{e_{ij}\},\cdots)$.  The total action amplitude $\ee^{S}$ for a
configuration $\{v_i\},\{e_{ij}\},\cdots$ (or a path) is given by
\begin{align}
\ee^{S}=
\prod_{(ij \cdots k)} T_{ij \cdots k}(\{v_i\},\{e_{ij}\},\cdots)
\end{align}
where $\prod_{(ij \cdots k)}$ is the product over all the $n$-cells $(ij \cdots
k)$.  Our lattice theory is defined by the following imaginary-time path
integral (or partition function)
\begin{align}
 Z(\{v^\text{bdry}_i, e^\text{bdry}_{ij},\cdots\}, C_N^{d+1},T)=
\hskip -1em
\sum_{ \{v_i\},\{e_{ij}\},\cdots }
\hskip -1em
\ee^{S} := |\Psi(C_N^{d+1})\>
\end{align}
where $\sum_{ \{v_i\},\{e_{ij}\},\cdots }$ only sum over the indices inside the
space-time complex, and the indices $v^\text{bdry}_i,
e^\text{bdry}_{ij},\cdots$ on the boundary of the space-time complex are fixed.
We see that on space-time with boundary, the path integral gives rise to a wave
function on the boundary $|\Psi\>$.  On space-time with no boundary, the path
integral gives rise to a complex number -- the partition function
$Z(C_N^{d+1},T)$.  (In the above dicussion, some important details are ignored.
More precise description can be found in the supplementary material and in
\Ref{KW1458}.)

In the $N\to \infty$ thermodynamic limit, the  partition function
is roughly given by
\begin{align}
\label{Ztop}
&\ \ \ \ Z(\{v^\text{bdry}_i, e^\text{bdry}_{ij}\},C^{d+1}_N,T) 
\nonumber \\
&=
\ee^{S^\text{eff}_N}Z^\text{top}(\{v^\text{bdry}_i, e^\text{bdry}_{ij}\},C^{d+1}_N,T),
\end{align}
where $S^\text{eff}_N=\int_\text{space-time} \text{energy-density} \propto N$,
and $Z^\text{top}(\{v^\text{bdry}_i, e^\text{bdry}_{ij}\},C^{d+1}_N,T)$ is
independent of $N$.  (The notion of topological partition function
$Z^\text{top}$ and topological path integral are discussed in more detail in
the supplementary material and in \Ref{KW1458}.) We see that the leading term
$S^\text{eff}_N$ behaves like a volume. Thus we will call $V(C^{d+1}_N, T)$
defined by 
\begin{align}
\label{Vdef}
V(C^{d+1}_N, T) = \log \sqrt{\<\Psi|\Psi\>}
\end{align}
as the (classical) space-time volume.  In other words, the many-body system
described by the tensor $T$ and triangulation $C^{d+1}_N$ give raise to a
definition of classical volume of the space-time.  At the leading order of $N$,
such a classical volume, $V(C^{d+1}_N, T)=S^\text{eff}_N=\int_\text{space-time} \text{energy-density}$, satisfies the inclusion-exclusion property: Let $Y$ be a
$d$-dimensional manifold with a Riemannian metric $\gamma$ and
$\mathcal{M}_{(Y,\gamma)}$ the set of all Riemannian manifolds $(X,g)$ such
that $\partial X=Y$ and $g|_Y=\gamma$.  Then the volume functional $V:
\mathcal{M}_{(Y,\gamma)}\rightarrow \mathbb{R}$ satisfies the
inclusion-exclusion formula 
\begin{align}
\label{cvadd}
 V(A)+V(B)= V(A\cup B)+V(A\cap B) , 
\end{align}

We like to mention that the Euler characteristic $\chi(X)$ of a topological
space $X$ usually appears in the path integrals as a prefactor $a^{\chi}$ (just
like $\ee^{S^\text{eff}_N}$) and behave like a volume.  The Euler characteristic has an
axiomatic characterization.  The Euler characteristic $\chi(X)$ is essentially
the only homotopy invariant function on all topological spaces that satisfies
the multiplicative property $\chi(X\times Y)=\chi(X)\chi(Y)$ and the
inclusion-exclusion formula $\chi(X)=\chi(A)+\chi(B)-\chi(A\cap B)$ if $X=A\cup
B$.  But there are no known axiomatic characterizations of the volume
functional.

To discuss volume more precisely, \ie to include both terms at the $N$-order
and the $N^0$-order, it is better to introduce a notion of quantum volume, or
q-volume.  A  q-volume is not a real number. It is a vector, \ie the wave
function $|\Psi\>$.  The norm of the q-volume gives rise to the corresponding
classical volume.  We like to stress that the above definition of q-volume is
very general. It applies to both gapped many-body systems and gapless many-body
systems.  In this paper, we will concentrate on gapped many-body systems.

A topological invariant that is closely related to volume is the Gromov norm of
the fundamental class of a manifold.  The Gromov norm behaves well with respect
to covering maps, so one test for quantum volume would be to study its behavior
under covering maps. Quantum volumes also satisfy some neutrality and gluing
properties.
For a many-body system described by tensor $T$, its q-volume satisfies
\begin{align}
\label{qvadd}
&\ \ \ \
|\Psi(C^{d+1}_1 \cup C^{d+1}_2)\>
\nonumber\\
&
=\Tr_{\prt C^{d+1}_1 \cap \prt C^{d+1}_2}  [|\Psi(C^{d+1}_1)\>\otimes |\Psi(C^{d+1}_2)\>] 
\end{align}
if the space-time complexes $C^{d+1}_1$ and $C^{d+1}_2$ only overlap on their
boundaries.  Here $\Tr_{\prt C^{d+1}_1 \cap \prt C^{d+1}_2}$ traces over the
degrees of freedom on the overlapped boundaries $\prt C^{d+1}_1 \cap \prt
C^{d+1}_2$.  Since $|\Psi\>$'s are exponentially large $\ee^{S^\text{eff}_N}$
in large $N$ limit, from \eqref{qvadd}, we can show that, in thermodynamic
limit, the corresponding classical volume satisfies
\begin{align}
\frac{ V(C^{d+1}_1 \cup C^{d+1}_2) }{ V(C^{d+1}_1) + V(C^{d+1}_2)}=  1
\end{align} 
which is a special case of \eqref{cvadd}.

~

\noindent
\textbf{Topological invariant through q-volume and surgery}:
In general the partition function $Z(C^{d+1}_N,T) =\ee^{S^\text{eff}_N}
Z^\text{top}$ of a many-body system is not a topological invariant even when
the many-body system described by the tensor $T$ realizes a topologically
ordered state.  In the absence of translation symmetry, it is not trivial to
separate the non-universal part $ \ee^{S^\text{eff}_N}$ from the topological invariant
$Z^\text{top}$ by just knowing $Z(C^{d+1}_N,T)$.

To achieve the separation, we note that, the $S^\text{eff}_N$ part of $\log
|Z(C^{d+1}_N,T)|$ is the standard classical volume of the space-time
that satisfy the inclusion-exclusion property \eqref{cvadd}.  Thus, we can
separate the $Z^\text{top}$ part  of the partition function, since
$Z^\text{top}$ violates the properties of classical volume.  $Z^\text{top}$ is
the topological invariant that reflects the non-trivial topological order in
the system.  In other words, for a system with trivial topological order, the
classical volume axioms are satisfied even at $\ee^{N^0}$ order, \ie
$Z^\text{top}=1$ (see \eqref{vRG}).

As an application of the above idea, we consider the following ratio of two
partition functions for a many-body system described by a tensor-set $T$:
\begin{align}
\label{tinv}
{
Z\Big( \bmm \includegraphics[height=0.6in]{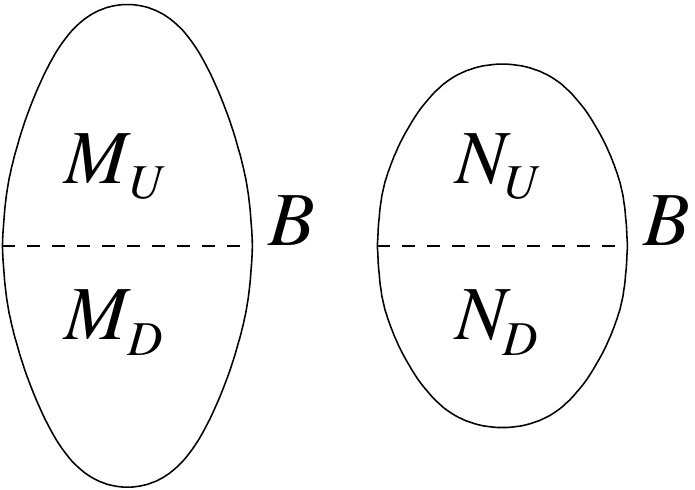} \emm, T\Big)
}\Big/{
Z\Big( \bmm \includegraphics[height=0.6in]{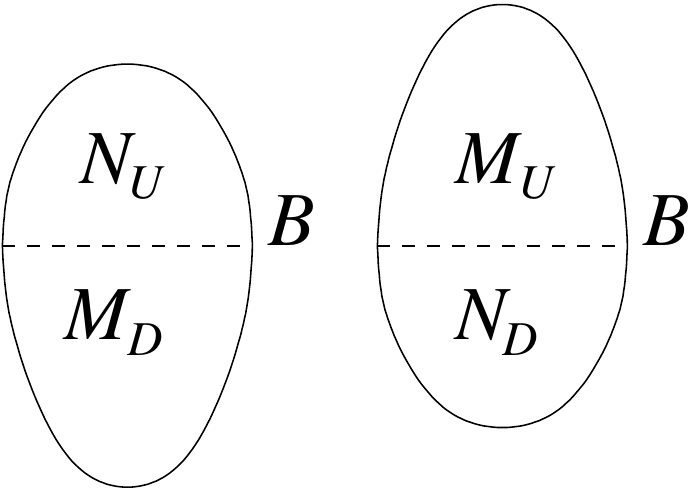} \emm, T\Big)
}
\end{align}
Here, we divide the $d+1$-dimensional space-time $M$ into two parts $M_U$ and
$M_D$ by a $d$-dimensional boundary with a triangulation $B$.  We also divide
the other space-time $N$ into two parts $N_U$ and $N_D$ by a boundary with the
same triangulation $B$.  This allows us to glue $M_U$ with $N_D$ and $N_U$ with
$M_D$.  

If $V=\log |Z|$ exactly satisfy the inclusion-exclusion property of the
classical volume, then the above ratio \eqref{tinv} will be 1.  However, in
general, the subleading $N^0$-term in $V=\log |Z|$ does not satisfy the
inclusion-exclusion property.  Such subleading terms will make the ratio
\eqref{tinv} to differ from 1.  But for a system $T$ with trivial topological
order, we find that the above ratio \eqref{tinv} will be 1 in the thermodynamic
limit.  This is because the partition functions and their ratio is invariant
under the tensor network renormalization transformations which coarse grain the
tensor network away from the boundary $B$.  If the tensor $T$ describes a
trivial topological order, the tensor network will flow to a corner-double-line
tensor network  in 1+1D or a similar structured tensor network in higher
dimensions \cite{GW0931}:  
\begin{align}
\label{vRG}
&\ \ \ \
{
Z\Big( \bmm \includegraphics[height=0.6in]{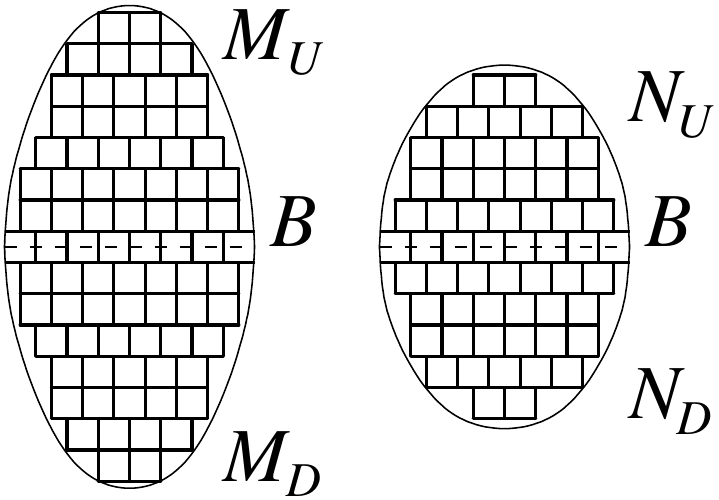} \emm, T\Big)
}\Big/{
Z\Big( \bmm \includegraphics[height=0.6in]{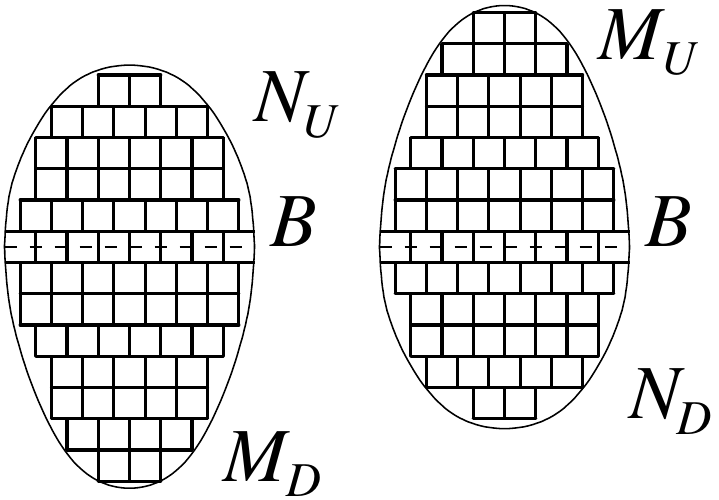} \emm, T\Big)
}
\\
&=
{
Z\Big( \bmm \includegraphics[height=0.6in]{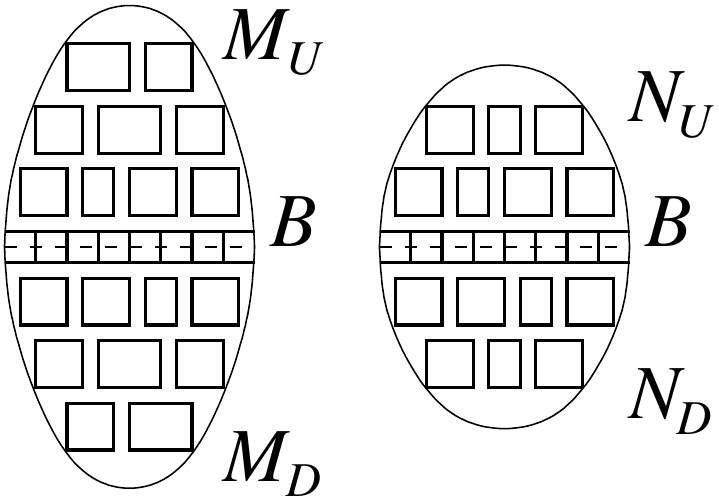} \emm, T\Big)
}\Big/{
Z\Big( \bmm \includegraphics[height=0.6in]{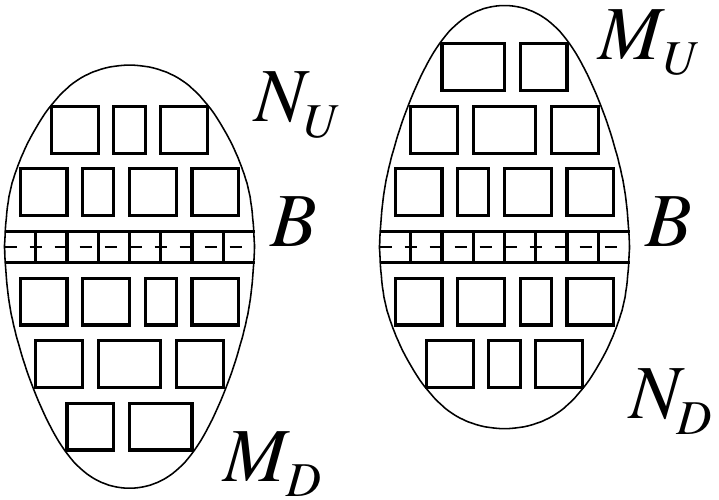} \emm, T\Big)
}=1
\nonumber 
\end{align}
This allows us to show the ratio
\eqref{tinv} to be 1, if the system has no topological order.  Thus the ratio
\eqref{tinv} is a topological invariant that can characterize non-trivial
topological orders in the system.  

The following ratio is also a topological invariant
\begin{align}
\label{ZMBM}
{
Z\Big( \bmm \includegraphics[scale=0.3]{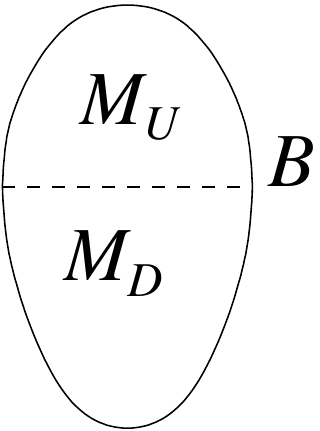} \emm, T\Big)
}\Big/\sqrt{
Z\Big( \bmm \includegraphics[scale=0.25]{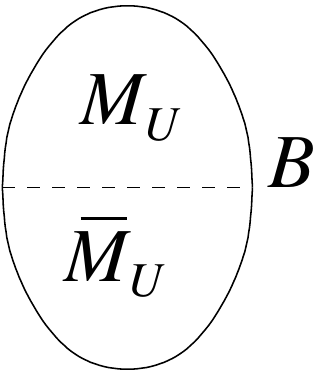} \emm, T\Big)
Z\Big( \bmm \includegraphics[scale=0.25]{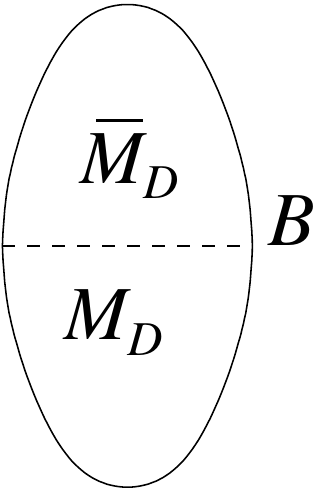} \emm, T\Big)
}
\end{align}
The above ratio is calculated by dividing the closed space-time $M$ into two
parts $M=M_U\cup M_D$.  It not only dependent on the space-time  $M$, it also
depends on $M_U$ and  $M_D$, \ie how we partition $M$.  Notice that the
space-time with boundary, $M_U$ and  $M_D$, give rise to two vectors
$\<\Psi(M_U)|$ and  $|\Psi(M_D)\>$, which are not normalized.  The above ratio
is simply the overlap of $\<\Psi(M_U)|$ and  $|\Psi(M_D)\>$ after
normalization:
$\frac{\<\Psi(M_U)|\Psi(M_D)\>
}{\sqrt{\<\Psi(M_U)|\Psi(M_U)\>}\sqrt{\<\Psi(M_D)|\Psi(M_D)\>}
}$.

Let us apply the above approaches to construct some
topological invariants.  First, for $d+1$D many-body systems with unique gapped
liquid ground state on $S^d$,
\begin{align}
&\ \ \ \
{
Z\Big( \bmm \includegraphics[height=0.6in]{MNB} \emm, T\Big)
}\Big/{
Z\Big( \bmm \includegraphics[height=0.6in]{MNB1} \emm, T\Big)
} 
\nonumber\\
&=
\frac{
 Z(M_U \cup M_D) Z(N_U \cup N_D) 
}{
 Z(N_U \cup M_D) Z(M_U \cup N_D) 
}
=\Big|_{B=S^d} 1
\end{align}
So when the partition boundary is a sphere $B=S^d$, the above ratio fails to
give rise to any non-trivial topological invariant.  Thus, the connected sum
decomposition does not give rise to non-trivial topological invariants.  The
non-trivial topological invariants may arise when the division has a
non-trivial cross section $B$ beyond a sphere.

One such topological invariant is obtained by choosing $M_U = D^2\times S^1,\
M_D = D^2\times S^1,\  N_U = S^1\times D^2,\  N_D = S^1\times D^2$.  We find
\begin{align}
\label{ZD2}
&\ \ \ \
\frac{
 Z(D^2\times S^1 \cup D^2\times S^1) Z(S^1\times D^2 \cup S^1\times D^2)
}{
 Z(S^1\times D^2 \cup D^2\times S^1) Z(D^2\times S^1 \cup S^1\times D^2)
}
\nonumber\\
&
=\Big|_{N\to\infty}
\Big( \frac{
Z^\text{top}(S^2\times S^1)
}{
Z^\text{top}(S^3)
} \Big)^2
= D^2
\end{align}
which allows us to calculate the total quantum dimension $D^2 =\sum d_i^2$ of a
2+1D topologically ordered state.  Here, $D^2\times S^1 \cup D^2\times S^1 =
S^2\times S^1$ is obtained by glueing two solid tori $D^2\times S^1$, and
$S^1\times D^2 \cup D^2\times S^1 = S^3$ is obtained by glueing two solid tori
in a twisted way.  

Also $N\to \infty$ is the limit of more and more refined triangulation of the
space-time (\ie the thermodynamics limit in condensed matter physics).  To
obtain the first equal sign in \eqref{ZD2} we have used the fact that the
leading term $S^\text{eff}_N$ in the partition function satisfies the inclusion-exclusion
property of the classical volume in $N\to \infty$ limit.  This is because the
leading $S^\text{eff}_N$ is given by the integration of local energy density over
space-time.  We can always tune the local energy density continuously without
encounter any phase transition.  Thus we can tune $S^\text{eff}_N$ to zero without any
phase transition.  This leads to the first equal sign in \eqref{ZD2}.

Another topological invariant is given by
\begin{widetext}
\begin{align}
&\ \ \ \
\frac{
 Z(D^2\times S^1\times S^1 \cup D^2\times S^1\times S^1) Z(S^1\times D^2\times S^1 \cup S^1\times D^2\times S^1)
}{
 Z(S^1\times D^2\times S^1 \cup D^2\times S^1\times S^1) Z(D^2\times S^1\times S^1 \cup S^1\times D^2\times S^1)
}
=\Big|_{N\to\infty}
\Big( \frac{
Z^\text{top}(S^2\times S^1\times S^1)
}{
Z^\text{top}(S^3\times S^1)
} \Big)^2
= N_p^2
\end{align}
\end{widetext}
which allows us to calculate the number $N_p$ of topological types of point-like
excitations of a 3+1D topologically ordered state.  

We can also use \eqref{ZMBM} to construct more topological invariants.  First,
let $M_U$ and $M_D$ be handlebodies of genus $g$, and let $f$ be an orientation
reversing homeomorphism from the boundary of $B=\prt M_U$ to the boundary of
$B=\prt M_D$. By gluing $M_U$ to $M_D$ along $B$ we obtain the compact oriented
3-manifold $M = V \cup_f W$.  Every closed, orientable three-manifold may be so
obtained, which is called a Heegaard splitting.  Thus we can construct a
topological invariant for each orientable three-manifold and its Heegaard
splitting.

More specifically, we can choose $M_U=D^2\times S^1$, $M_D=D^2\times S^1$, and
$f$ be a mapping from $S^1\times S^1$ to $S^1\times S^1$.  Thus $f$ is an
element in $SL(2,\Z)$.  In this case we find that
\begin{align}
\label{M11}
&
{
Z\Big( \bmm \includegraphics[scale=0.3]{MUD} \emm, T\Big)
}\Big/\sqrt{
Z\Big( \bmm \includegraphics[scale=0.25]{MUU} \emm, T\Big)
Z\Big( \bmm \includegraphics[scale=0.25]{MDD} \emm, T\Big)
}
=\Big |_{N\to \infty} 
\nonumber\\
&
{
Z^\text{top}\Big( \bmm \includegraphics[scale=0.3]{MUD} \emm, T\Big)
}\Big/\sqrt{
Z^\text{top}\Big( \bmm \includegraphics[scale=0.25]{MUU} \emm, T\Big)
Z^\text{top}\Big( \bmm \includegraphics[scale=0.25]{MDD} \emm, T\Big)
}
\nonumber\\
&=
(M_f)_{11},
\end{align}
where $M_f$ is the representation of $SL(2,\Z)$ in the quasiparticle basis
\cite{W9039,KW9327,ZGT1251}.

We may also choose $M_U=I^1\times T^2 \cup_f I^1\times T^2$, and $M_D=I^1\times
T^2$.  Note that $I^1\times T^2$ has two $T^2=S^1\times S^1$ boundaries.  $M_U$
is formed  by two  $I^1\times T^2$'s glued along one of the $T^2$ boundary with
a $f$-twist.  Then, we glue the two $T^2$ boundaries of $M_U$ and two  $T^2$
boundaries of $M_D$ directly without twist to form the total space-time
lattice.  In this case we find that
\begin{align}
\label{TrM}
&
{
Z\Big( \bmm \includegraphics[scale=0.3]{MUD} \emm, T\Big)
}\Big/\sqrt{
Z\Big( \bmm \includegraphics[scale=0.25]{MUU} \emm, T\Big)
Z\Big( \bmm \includegraphics[scale=0.25]{MDD} \emm, T\Big)
}
=\Big |_{N\to \infty} 
\nonumber\\
&
{
Z^\text{top}\Big( \bmm \includegraphics[scale=0.3]{MUD} \emm, T\Big)
}\Big/\sqrt{
Z^\text{top}\Big( \bmm \includegraphics[scale=0.25]{MUU} \emm, T\Big)
Z^\text{top}\Big( \bmm \includegraphics[scale=0.25]{MDD} \emm, T\Big)
}
\nonumber\\
&=
\Tr(M_f)/\text{dim}{M_f}.
\end{align}
$M_f$ is an important topological invariant that characterizes the
topological order in the many-body system.

The above expression allows us to compute the topological invariants
$(M_f)_{11}$ and $\Tr(M_f)$ using generic non-fixed point path integral $Z\Big(
\bmm \includegraphics[scale=0.2]{MUD} \emm, T\Big)$ in the thermodynamical
limit.  We like to stress that in \eqref{M11}, we need to choose the
triangulation on $M_U=D^2\times S^1$ and $M_D=D^2\times S^1$, such that the
induced triangulation on the common boundary $B=\prt D^2\times S^1$ is related
by the mapping $f$. 

We know that $SL(2,\Z)$ is generated by $S$ for $f=\bpm 0 & -1\\ 1 & 0 \\ \epm$
and $T$ for $f=\bpm 1 & 1\\ 0 & 1 \\ \epm$. By choosing different $f$'s, we can
obtain $S_{11}$, $T_{11}$, $(ST)_{11}$, \etc in the quasiparticles basis.

~

\noindent
\textbf{Summary}:
We introduce a notion of quantum volume for quantum many-body systems defined
on space-time lattice.  The quantum volume is not a positive number but a
vector in a Hilbert space, which satisfies an additive property \eqref{qvadd}.
We show that the norm of the quantum volume gives rise to classical volume that
satisfies the inclusion-exclusion property \eqref{cvadd} in the thermodynamic
limit.

For a many-body system with topological order, its partition function is not
universal and depends in the details of interaction.  Using the idea of quantum
volume and classical volume, we show how to compute topological invariants from
non-universal partition functions.  In particular, we show how to compute the
trace and the $(1,1)$ matrix element of the modular representation in
quasiparticle basis from non-universal 2+1D partition functions.

~

X.-G Wen is partially supported by NSF grant DMR-1506475, DMS-1664412, and
NSFC 11274192.  Z. Wang is partially funded by NSF grant DMS-1411212 and
FRG-1664351.

\vfill \eject

\appendix

\centerline{\large \bf
Supplementary Materials
}

\section{Many-body systems and path integral on space-time lattice } 
\label{path}

In this section, we will define many-body systems without translation symmetry
via space-time path integral.  We will define space-time path integral using
uniform tensors on arbitrary random space-time lattice.  Despite the tensors
are uniform, the random space-time lattice breaks the translation symmetry.
Later we will use such space-time path integral to define the quantum and
classical volumes of the random space-time lattice.

\subsection{Space-time complex}
\label{stcomp}

To define a Many-body system through a space-time path integral, we first
triangulate the $n$-dimensional space-time to obtain a simplicial complex
$C_N$ (see Fig. \ref{comp}).  Here we assume that all simplicial complexes are of bounded geometry 
in the sense that the number of edges that
connect to one vertex is bounded by a fixed value.  Also the number of
triangles that connect to one edge is bounded by a fixed value, \etc.

\begin{figure}[tb]
\begin{center}
\includegraphics[scale=0.6]{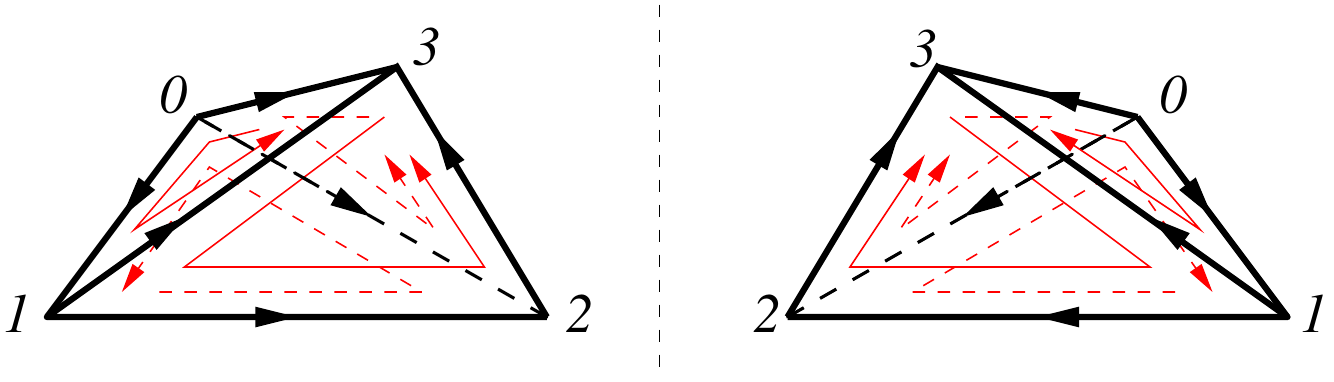} \end{center}
%Fig. 2
\caption{ (Color online) Two branched simplices with opposite orientations.
(a) A branched simplex with positive orientation and (b) a branched simplex
with negative orientation.  }
\label{mir}
\end{figure}

In order to define a generic lattice
theory on the space-time complex $C_N$, it is important to give the
vertices of each simplex a local order.  A nice local scheme to order  the
vertices is given by a branching structure.\cite{C0527,CGL1172,CGL1204} A
branching structure is a choice of orientation of each edge in the
$n$-dimensional complex so that there is no oriented loop on any triangle (see
Fig. \ref{mir}).

The branching structure induces a \emph{local order} of the vertices on each
simplex.  The first vertex of a simplex is the vertex with no incoming edges,
and the second vertex is the vertex with only one incoming edge, \etc.  So the
simplex in  Fig. \ref{mir}a has the following vertex ordering:
$0<1<2<3$.

The branching structure also gives the simplex (and its sub simplexes) an
orientation denoted by $s_{ij \cdots k}=1,*$.  Fig. \ref{mir} illustrates two
$3$-simplices with opposite orientations $s_{0123}=1$ and $s_{0123}=*$.  The
red arrows indicate the orientations of the $2$-simplices which are the
subsimplices of the $3$-simplices.  The black arrows on the edges indicate the
orientations of the $1$-simplices.

\subsection{Path integral on a space-time complex}
\label{stpath}

\begin{figure}[tb]
\begin{center}
\includegraphics[scale=0.6]{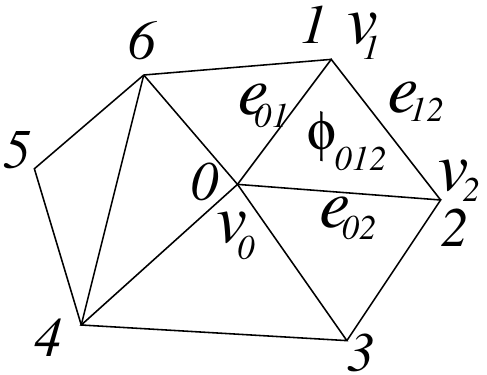} \end{center}
%Fig. 1
\caption{ 
A 2-dimensional complex. 
The vertices (0-simplices) are labeled by $i$.
The edges (1-simplices) are labeled by $\<ij\>$.
The faces (2-simplices) are labeled by $\<ijk\>$.
The degrees of freedoms may live on 
the vertices (labeled by $v_i$),
on the edges (labeled by $e_{ij}$)
and on the faces (labeled by $\phi_{ijk}$).
}
\label{comp}
\end{figure}

The degrees of freedom of our lattice model live on the vertices  (denoted by
$v_i$ where $i$ labels the vertices), on the edges (denoted by $e_{ij}$ where
$\<ij\>$ labels the edges), and on other high dimensional simplicies of the space-time
complex (see Fig. \ref{comp}).  The action amplitude $\ee^{S_\text{cell}}$ for an $n$-cell $(ij
\cdots k)$ is complex function of $v_i$, $e_{ij},\cdots$: $T_{ij \cdots
k}(\{v_i\},\{e_{ij}\},\cdots)$.  The total action amplitude $\ee^{S}$ for
a configuration $\{v_i\},\{e_{ij}\},\cdots$ (or a path) is given by
\begin{align}
\label{eS}
\ee^{S}=
\prod_{(ij \cdots k)} [T_{ij \cdots k}(\{v_i\},\{e_{ij}\},\cdots)]^{s_{ij \cdots k}}
\end{align}
where $\prod_{(ij \cdots k)}$ is the product over all the $n$-cells $(ij \cdots
k)$.  Note that the contribution from an $n$-cell $(ij \cdots k)$ is
$T_{ij \cdots k}(\{v_i\},\{e_{ij}\},\cdots)$ or $T^*_{ij \cdots k}(\{v_i\},\{e_{ij}\},\cdots)$
depending on the orientation $s_{ij \cdots k}$ of the cell.
Our lattice theory is defined
by the following imaginary-time path integral (or partition function)
\begin{align}
\label{Zpath}
 Z=\sum_{ \{v_i\},\{e_{ij}\},\cdots }
\prod_{(ij \cdots k)} [T_{ij \cdots k}(\{v_i\},\{e_{ij}\},\cdots)]^{s_{ij \cdots k}}.
\end{align}
Clearly, the partition function $Z$ depends on the space-time $M$, so we denote
it as $Z(M)$.  It is also clear that the  partition function on a disjoint
union of $M$ and $N$ is given by the product of the two partition functions on
$M$ and on $N$:
\begin{align}
 Z(M \sqcup N)=Z(M)Z(N)
\end{align}

We would like to point out that, in general, the path integral may also depend
on some additional weighting factors $w_{v_i}$, $\tAw{d}ij$, \etc
(see \eqref{Z3d}).  In this section, for simplicity, we will
assume those  weighting factors are all equal to $1$.

In the above path integral \eq{Zpath}, we have assigned the same action
amplitude $T_{ij \cdots k}(\{v_i\},\{e_{ij}\},\cdots)$ to each simplex $(ij
\cdots k)$.  Such a  path integral is called a \emph{uniform path integral}.
For simplicity, in this paper, we only study systems described by uniform path
integral.  But our discussion also apply the more complicated cases where
different  simplices $(ij \cdots k)$ have different action amplitudes.

\begin{figure}[tb]
\begin{center}
\includegraphics[scale=0.5]{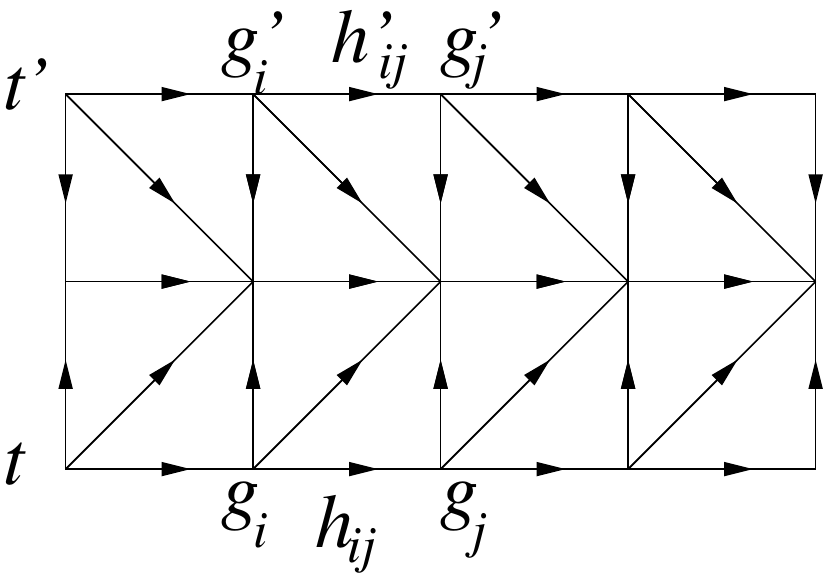} \end{center}
%Fig. 3
\caption{
Each time-step of evolution is given by the path integral on a particular form
of branched graph.  Here is an example in 1+1D. 
}
\label{tStep}
\end{figure}

\begin{figure}[tb]
\begin{center}
\includegraphics[scale=0.43]{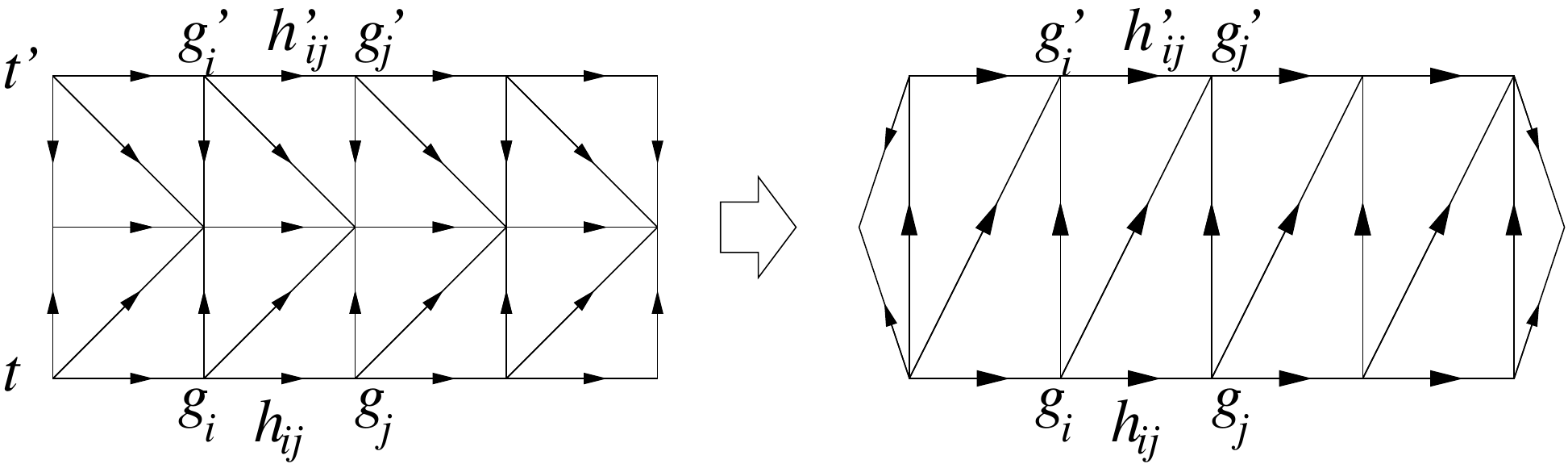} \end{center}
%Fig. 4
\caption{
The reduction of double-layer time-step to single-layer time-step on space with
boundary for an 1+1D topological path integral.
}
\label{stStep}
\end{figure}

\subsection{Path integral on a space-time complex with boundary}

In the last subsection, we have defined the path integral on a space-time
complex without boundary.  In this case, all the indices on vertices, edges,
\etc are summed over.  The resulting partition function $Z$ is just a complex
number.

If the space-time manifold has a boundary $\prt M^{d+1} =M_B^d$, then the
triangulation $C^{d+1}_N$ of $M^{d+1}$ has the following property: all the
vertices in $C^{d+1}_N$ that are on the boundary $M_B^d$ form a subcomplex
$B^d_{N'}$, such that $B^d_{N'}$ is a triangulation of $M_B^d$.  In this
case, we say that the complex $C^{d+1}_N$ has a boundary which is given by
$B^d_{N'} \equiv \prt C^{d+1}_N$.

The path integral on $C^{d+1}_N$ with a boundary is defined differently: we
only sum over the indices on vertices, edges, \etc that are not on the boundary
$B^d_{N'}$.  The indices on the boundary $B^d_{N'}$ are fixed.  So the
resulting partition function $Z$ is a function of the indices on the boundary
$B^d_{N'}$.  We see that the boundary $B^d_{N'}$ gives rise to a Hilbert
space $\cH_{B^d_{N'}}$ formed by all the complex functions of the indices on
the boundary $B^d_{N'}$.  The partition function $Z$ on $C^{d+1}_N$ is a
vector in  $\cH_{B^d_{N'}}$ (\ie a particular complex function of the indices
on $B^d_{N'}$).  This is consistent with the Atiyah's definition of
topological quantum field theory \cite{A8875}.

\subsection{Path integral and Hamiltonian}
\label{pathham}

Consider a space-time complex of topology $M_\text{space}\times I$ where
$I=[t,t']$ represents the time dimension and $M_\text{space}$ is a closed space
complex (see Fig. \ref{tStep}).  The space-time complex $M_\text{space}\times
I$ has two boundaries: one at time $t$ and another at time $t'$.  A path
integral on the space-time complex $M_\text{space}\times I$ is a function of
the indices on the two boundaries, which give us an amplitude $Z[\{ v_i',
e_{ij}',\cdots  \}, \{ v_i, e_{ij},\cdots \}]$ from a configuration $\{ v_i,
e_{ij},\cdots \}$ at $t$ to another configuration $\{ v_i', e_{ij}',\cdots  \}$
at $t'$.  Here, $\{ v_i, e_{ij},\cdots \}$ and $\{ v_i', e_{ij}',\cdots  \}$
are the degrees of freedom on the boundaries (see Fig. \ref{tStep}).  We like
to interpret $Z[\{ v_i', e_{ij}',\cdots  \}, \{ v_i, e_{ij},\cdots  \}]$ as the
amplitude of an evolution in imaginary time by a Hamiltonian:
\begin{align}
& \ \ \ \
 Z[\{ v_i', e_{ij}',\cdots  \}, \{ v_i, e_{ij},\cdots \}] 
\nonumber\\
&=\< v_i', e_{ij}',\cdots  | \ee^{-(t'-t)H} |
v_i, e_{ij},\cdots  \> .
\end{align}
However, such an interpretation may not be valid since $ Z[\{ v_i',
e_{ij}',\cdots  \}, \{ v_i, e_{ij},\cdots \}]$ may not give raise to a
Hermitian matrix.  It is a worrisome realization that path integral and
Hamiltonian evolution may not be directly related.

Here we would like to use the fact that the path integral that we are
considering are defined on the branched graphs with a ``reflection'' property
(see \eq{eS}). We like to show that such path integral are better related
Hamiltonian evolution.  The key is to require that each time-step of evolution
is given by  branched graphs of the form in Fig. \ref{tStep}.  One can show
that $Z[\{ v_i', e_{ij}',\cdots  \}, \{ v_i, e_{ij},\cdots \}]$ obtained by summing over all in
the internal indices in the  branched graphs Fig. \ref{tStep}
has a form
\begin{align}
&\ \ \ \
 Z[\{ v_i', e_{ij}',\cdots  \}, \{ v_i, e_{ij},\cdots \}]
\\
&=\sum_{\{ v_i'', e_{ij}'',\cdots  \}}
 U^*[\{ v_i'', e_{ij}'',\cdots  \}, \{ v_i', e_{ij}',\cdots \}]
\nonumber\\
&\ \ \ \ \ \ \ \ \ \ \ \ \ \ \ \ \
 U[\{ v_i'', e_{ij}'',\cdots  \}, \{ v_i, e_{ij},\cdots \}]
\nonumber
\end{align}
and represents a positive-definite Hermitian matrix.  Thus the path integral of
the form \eq{eS} always correspond to a Hamiltonian evolution in imaginary
time.  In fact, the above $Z[\{ v_i', e_{ij}',\cdots  \}, \{ v_i, e_{ij},\cdots
\}]$ can be viewed as an imaginary-time evolution $T=\ee^{-\Del \tau H}$ for a
single time step.

\section{Topological path integral}
\label{toppath}

In this section, we will review some results from \Ref{KW1458}.

\subsection{Topological path integral and topological orders with gappable boundary}

The ground states of some many-body systems can have a special properties that
the ground states on systems with different size only different a stacking of a
product state, up to a local unitary transformation:
\begin{align}
 |\Psi_{N_2}\> = U_{LU} |\Psi_{N_1}\>\otimes |\Psi_{PS}\>
\end{align}
where $N_2>N_1$ describe the system size, $|\Psi_{PS}\>$ is a product state for
a system of size $N_2-N_1$, and $ U_{LU}$ is a local unitary transformation.
Such kind of ground states are called gapped liquid states
\cite{ZW1490,SM1403}. The gapped liquid states formally define the
topologically ordered states \cite{W9039,WN9077}.  

As many-body systems, all topologically ordered states are described by
path-integrals, and a path-integral can be described by a TN with finite
dimensional tensors on a space-time lattice (\ie a space-time complex).  Even
though topologically ordered states are all gapped, only some of them can be
described by the so called fixed-point path-integrals which are called
\emph{topological path integrals}:
\begin{defn} \textbf{Topological path integral} \\
(1) A topological path integral has an action 
amplitude that can be described by 
a TN with \emph{finite dimensional} tensors.\\
(2)  It is a sum of the action amplitudes for all the paths.
(The summation corresponds to the tensor contraction.) \\ 
(3)  Such a sum (called the partition function $Z^\text{top}(M)$) on a closed space-time
$M$ only depend on the topology of the space-time.  The partition function is
invariant under the local deformations and reconnections of the TN.\\
%(4)  To describe a local bosonic Lagrangian system, we 
%require the partition function to be well defined
%on space-time with any topology.\\
%(5)  To describe a local Hamiltonian qubit system, we only
%require the partition function to be well defined
%on space-time which is a mapping torus.
\end{defn}\noindent
In the next section, we will give concrete examples of the topological
path integrals. The topological path integrals are closely related to 
topological orders with gappable boundary
 \cite{TV9265,LWstrnet,kirillov,balsam-kirillov}. We like to conjecture that \cite{KW1458} 
\begin{conj} 
All topological orders with gapped boundary
are described by topological path integrals.
\end{conj} \noindent
We make such a conjecture because we believe that the tensor network
representation that we are going to discuss is the most general one.  It can
capture all possible fixed-point tensors \cite{CGW1038} under renormalization
flow generated by the coarse-graining of the TN \cite{VC0466,LN0701}, and those
fixed-point tensors give rise to  topological path integrals.  

We also like to remark that we cannot say that all topological path integrals
describe  topological orders with gapped boundary, since some  topological path
integrals are stable while others are unstable (which means a small
perturbation of the tensors will result in a different fixed-point tensor under
renormalization flow).  Only the stable topological path integrals
describe topological order.  Here we like to conjecture that \cite{KW1458}
\begin{conj} 
\label{Zstable}
A topological path integral in $(d+1)$-dimensional space-time constructed with
finite dimensional tensors is stable iff the partition function of the
topological path integral satisfies $|Z^\text{top}(S^1\times S^d)|=1$.
\end{conj}\noindent
Note that $Z^\text{top}(S^1\times S^d)$ is the ground state degeneracy on
$d$-dimensional space $S^d$.  If a system has a gap and the ground degeneracy
is 1, a small perturbation cannot do much to destabilize the state.  So
$Z^\text{top}(S^1\times S^d)=1$ is the sufficient condition for a stable  topological path
integral.  This argument implies that if the ground degeneracy is 1 on $S^d$,
then the system has no locally distinguishable ground state, and the ground
state degeneracy on space with other topologies are all robust against any
small perturbations.  

Since the topological path integrals are independent of re-triangulation of the
space-time, the partition function on a closed space-time only depends on the
topology of the space-time.
We like to point out that two topological path integrals, $Z^\text{top}(M)$ and
$\t Z^\text{top}(M)$, can be smoothly connected if the two topological path
integrals differ by 
\begin{align}
\label{eupon}
 \t Z^\text{top}(M)/Z^\text{top}(M)=W^{\chi(M)}  
\ee^{\ii \sum_{\{n_i\}} \phi_{n_1n_2\cdots} \int_M P_{n_1n_2\cdots}} ,
\end{align}
where $\chi(M)$ is the Euler number of $M$ and $P_{n_1n_2\cdots}$ are
combinations of Pontryagin classes: $P_{n_1n_2\cdots}=p_{n_1}\wedge
p_{n_2}\wedge \cdots$ on $M$.  $Z^\text{top}(M)$ and $\t Z^\text{top}(M)$ are
connected since complex numbers $W$ and $\phi_{n_1n_2\cdots}$ are not
quantized.  

Eqn. (\ref{eupon}) may be the only local topological invariant
that is not quantized (\ie $W$ and $\phi_{n_1n_2\cdots}$
can be any complex numbers). Thus \cite{KW1458}
\begin{conj} 
\label{ZZpEq}
 $Z^\text{top}(M)$ and $\t Z^\text{top}(M)$ are connected iff they are related by \eqn{eupon}.
\end{conj}\noindent
In other words, if two  topological path integrals produce two
topology-dependent partition functions that differ by a factor $W^{\chi(M)}
\ee^{\ii \sum_{\{n_i\}} \phi_{n_1n_2\cdots} \int_M P_{n_1n_2\cdots}}$, then the
two topological path integrals describe the same topological order.

\begin{figure}[tb]
\begin{center}
\includegraphics[scale=0.6]{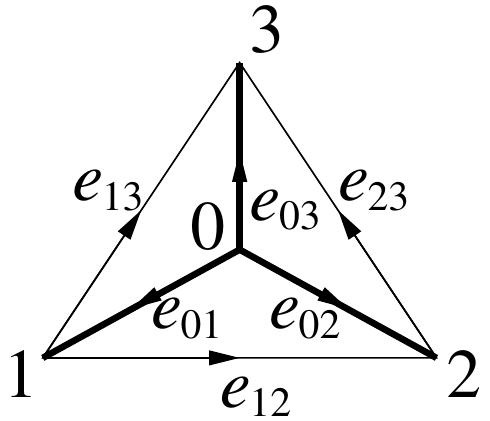}
%Fig. 5
\end{center}
\caption{
The tensor $\tC{C}0123$ is associated with a tetrahedron, which has a branching
structure.  If the vertex-0 is above the triangle-123, then the tetrahedron
will have an orientation $s_{0123}=*$.  If the vertex-0 is below the
triangle-123, the tetrahedron will have an orientation $s_{0123}=1$. The
branching structure gives the vertices a local order: the $i^{th}$ vertex has
$i$ incoming edges.  
}
\label{tetr}
\end{figure}

Summarizing the above discussions:\\
(1) All topological orders with gappable
boundary are described by \emph{stable} topological path integral constructed
with finite dimensional tensors.\\
(2) All stable topological path integrals describe topological orders with gappable
boundary.\\
(3) All stable topological path integrals
related by \eqn{eupon} describe the same topological order.
\\
So, we may view the stable topological path integrals as a 
classification of topological orders with gappable
boundary.

\subsection{Examples of topological path integrals in 2+1D}

\begin{figure}[tb]
\begin{center}
\includegraphics[scale=0.5]{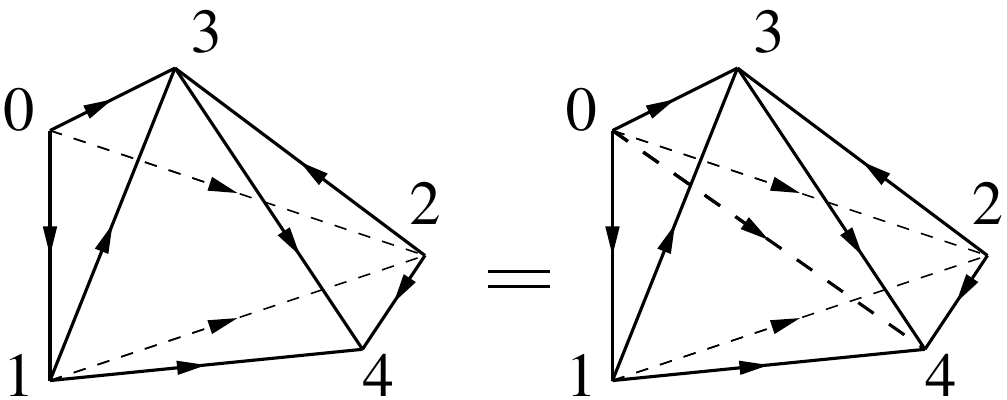}
%Fig. 6
\end{center}
\caption{
A retriangulation of a 3D complex.
}
\label{2to3}
\end{figure}
\begin{figure}[tb]
\begin{center}
\includegraphics[scale=0.5]{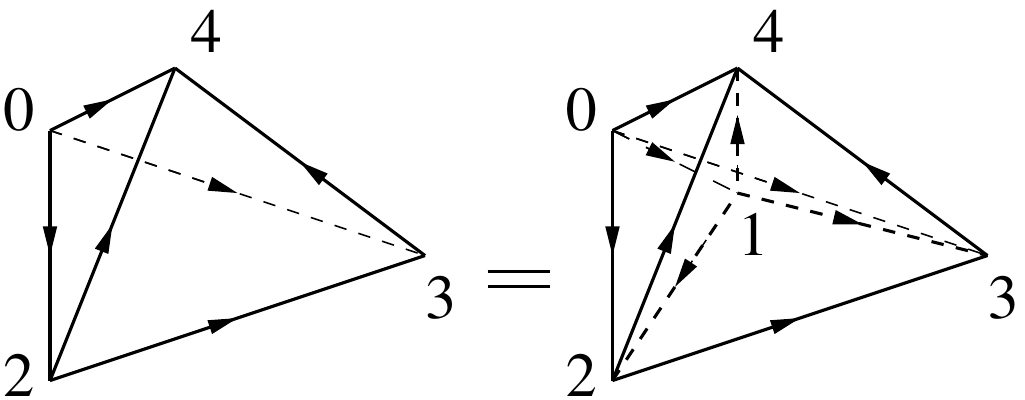}
%Fig. 7
\end{center}
\caption{
A retriangulation of another 3D complex.
}
\label{1to4}
\end{figure}

The topological path integral that describes a 2+1D topologically ordered state
with a gapped boundary can be constructed from a tensor set $T$ of two real
and one complex tensors: $T=(w_{v_0}, \tAw{d}01,\tC{C}0123)$.  The complex
tensor $\tC{C}0123$ can be associated with a tetrahedron, which has a
branching structure (see Fig.  \ref{tetr}).  A branching structure is a choice
of orientation of each edge in the complex so that there is no oriented loop on
any triangle (see Fig.  \ref{tetr}).  Here the $v_0$ index is associated with
the vertex-0, the $e_{01}$ index is associated with the edge-$01$, and the
$\phi_{012}$ index is associated with the triangle-$012$.  They represents the
degrees of freedom on the vertices, edges, and the triangles.

Using the tensors, we can define the topological path integral on any 3-complex
that has no boundary:
\begin{align}
\label{Z3d}
 Z=\sum_{ v_0,\cdots; e_{01},\cdots; \phi_{012},\cdots}
&\prod_\text{vertex} w_{v_{0}} 
\prod_\text{edge} \tAw{d}01\times
\\
&
\prod_\text{tetra} [\tC{C}0123 ]^{s_{0123}}
\nonumber 
\end{align}
where $\sum_{v_0,\cdots; e_{01},\cdots; \phi_{012},\cdots}$ sums over all the
vertex indices, the edge indices, and the face indices, 
$s_{0123}=1$ or $*$
depending on the orientation of tetrahedron (see Fig.  \ref{tetr}).
We want to choose the tensors $(w_{v_0}$, $\tAw{d}01$, $\tC{C}0123)$ such that
the path integral is re-triangulation invariant.  Such a topological path
integral describes a topologically ordered state in 3-space-time
dimensions and also define an topological order with gappable boundary.

On the complex $C^3_N$ with boundary: $B^2_{N'}= \prt C^3_N$,
the partition function is defined differently:
\begin{align}
\label{Z3dB}
 Z=\sum_{ \{ v_i; e_{ij}; \phi_{ijk} \} }
&\prod_{\text{vertex}\notin B^2_{N'}} w_{v_{0}} 
\prod_{\text{edge}\notin B^2_{N'}} \tAw{d}01\times
\\
&
\prod_\text{tetra} [\tC{C}0123 ]^{s_{0123}}
\nonumber 
\end{align}
where $\sum_{v_i; e_{ij}; \phi_{ijk}}$ only sums over the vertex indices, the
edge indices, and the face indices that are not on the boundary.  The resulting
$Z$ is actually a complex function of $v_{i}$'s, $e_{ij}$'s, and $\phi_{ijk}$'s
on the boundary $B^2_{N'}$: $Z=Z(\{v_{i};e_{ij};\phi_{ijk}\})$.  Such a
function is a vector in $\cH_{B^2_{N'}}$.  We will denote such a vector as
$|\Psi(C^3_N)\>$.

We also note that the vertices and the edges are attached with the tensors
$w_{v_{i}}$ and $\tAw{d}01$.  But when we glue two boundaries together, those
tensors $w_{v_{i}}$ and $\tAw{d}ij$ are added back.  So the tensors $w_{v_{i}}$
and $\tAw{d}ij$ defines the inner product in the boundary Hilbert space
$\cH_{B^2_{N'}}$.
Therefore, we require $w_{v_{i}}$
and $\tAw{d}ij$ to satisfy the following unitary condition
\begin{align}
 w_{v_{i}} > 0, \ \ \ \tAw{d}ij >0.
\end{align}

\begin{widetext}
The invariance of $Z$ under the re-triangulation in Fig. \ref{2to3}
requires that
\begin{align}
\label{CC23}
&\ \ \
\sum_{\phi_{123}} \tC{C}0123 \tC{C}1234
\nonumber\\
&=
\sum_{e_{04}} \tAw{d}04
\sum_{ \phi_{014} \phi_{024} \phi_{034} }
\tC{C}0124 
\tC{C^*}0134
\tC{C}0234 .
\end{align}
We would like to mention that there are other similar conditions for different
choices of the branching structures.  The branching structure of a tetrahedron
affects the labeling of the vertices.

The invariance of $Z$ under the re-triangulation in Fig. \ref{1to4}
requires that
\begin{align}
\label{CC14}
&
\tC{C}0234
=
\sum_{e_{01}e_{12}e_{13}e_{14},v_1} w_{v_1} 
\tAw{d}01 \tAw{d}12 \tAw{d}13 \tAw{d}14 
\sum_{ 
\phi_{012} 
\phi_{013} 
\phi_{014} 
\phi_{123} 
\phi_{124} 
\phi_{134} 
}
\\
&\ \ \ \ \ \ \ \ \ \ \ \ 
\tC{C}0123
\tC{C^*}0124 
\tC{C}0134
\tC{C}1234
\nonumber 
\end{align}
Again there are other similar conditions for different choices of the branching
structures.
\end{widetext}

The above two types of the conditions are sufficient for producing a
topologically invariant partition function $Z$, which is nothing but the
topological invariant for three manifolds introduced by Turaev and
Viro.\cite{TV9265} Again, two different solutions are regarded as equivalent if
they produces the same topology-dependent partition function 
for any closed space-time.

It is also clear that the above construction of topological path integrals can
be easily generalized to any other dimensions. This gives rise to a
classification of topological orders with gappable boundary in higher
dimensions.

\section{Quantum volume and its property}

We have seen that when a tensor set, for example $T=(w_{v_0},
\tAw{d}01,\tC{C}0123)$, satisfy the conditions \eqref{CC23} and \eqref{CC14},
its path integral on different space-time complexes will describe the same
topological phase.  If we change the tensors by a small amount, the tensor set
will still describe the same topological phase for different space-time
complexes.  With such a more careful definition of path integral in term of the
tensor set and the space-time complex $C_N^{d+1}$ with branching structure, we
can define quantum volume more precisely.

For example, on 2+1D space-time complex $C^3_N$ with boundary $B^2_{N'}$, the
path integral produces a complex function $Z(\{v_i;e_{ij};\phi_{ijk}\})$ with
$\{v_i;e_{ij};\phi_{ijk}\} \subset B^2_{N'}$, which is a vector $|\Psi(C^3_N)\>
\in \cH_{B^2_{N'}}$ (see \eqref{Z3dB}).  
The inner
product in $\cH_{B^2_{N'}}$ is defined through the weight-tensors $w_{v_i}$
$\tAw{d}ij$:
\begin{align}
 \<\Psi|\Psi\>=\sum_{\{v_i;e_{ij};\phi_{ijk}\}}
\prod_i w_{v_i} \prod_{\<ij\>} \tAw{d}ij
|Z(\{v_i;e_{ij};\phi_{ijk}\})|^2 .
\end{align}
In this case the classical volume of $C^3_N$ is given by
\begin{align}
 V(C^3_N,T) = \log \sqrt{\<\Psi(C^3_N,T)|\Psi(C^3_N,T)\>}.
\end{align}
We can show that,
for a many-body system described by tensor set $T$,
its q-volume satisfies
\begin{align}
\label{qvadd1}
&\ \ \ \
|\Psi(C^{d+1}_1 \cup C^{d+1}_2)\>
\nonumber\\
&
=\Tr_{\prt C^{d+1}_1 \cap \prt C^{d+1}_2}  [|\Psi(C^{d+1}_1)\>\otimes |\Psi(C^{d+1}_2)\>] 
\end{align}
if the space-time complexes $C^{d+1}_1$ and $C^{d+1}_2$ only overlap on their
boundaries.  Here $\Tr_{\prt C^{d+1}_1 \cap \prt C^{d+1}_2}$ traces over the
\emph{internal} degrees of freedom on the overlapped boundaries $\prt C^{d+1}_1
\cap \prt C^{d+1}_2$ with the weighting tensors $w_{v_1}$,
$d^{v_1v_2}_{e_{12}}$, \etc for the \emph{internal} simplices on the overlapped
boundaries.  In other words, we only traces over the indices for the  simplices
\emph{inside} the overlapped boundaries (not those for the  simplices on the
boundary of the overlapped boundaries).  The summation of each index is
weighted by the corresponding weight tensor $w_{v_1}$ or $d^{v_1v_2}_{e_{12}}$
\etc.  \eqref{qvadd1} is a key property of the quantum volume. 

\bibliography{../../bib/wencross,../../bib/all,../../bib/publst,./local} 

\end{document}